\documentclass[reprint,longbibliography,pre,amsmath,amssymb,amsfont,url]{revtex4-2}
\usepackage{bold-extra}
\usepackage{amsthm}

\usepackage[colorlinks, allcolors=blue]{hyperref}
\usepackage{graphicx} 
\usepackage{csquotes}
\usepackage{mathtools}
\usepackage{enumitem}
\usepackage{booktabs}

\newtheorem*{thm}{Theorem}

\newcommand{\myquote}[1]{\textit{\textquote{#1}}}

\begin{document}
	\author{D. Lairez}
	\email{didier.lairez@polytechnique.edu}
	\affiliation{Laboratoire des solides irradi\'es, \'Ecole polytechnique, CEA, CNRS, IPP,
		%Institut Polytechnique de Paris, 
		91128 Palaiseau, France}
	\title{The fundamental difference between logical and thermodynamic irreversibilities, or, Why Landauer's result cannot be a physical principle}
	\date{\today}
	
	\begin{abstract}
		
	Landauer's \textquote{principle} claims that erasing one bit of information necessarily dissipates at least $Tln2$ of heat into the surroundings, making a possibly logically irreversible operation also thermodynamically irreversible. It is commonly accepted that this result is a fundamental principle of physics that definitively establishes the link between information and energy. 	
Here we show that this result cannot be general. In fact it
 comes: 1)~from a confusion between logical and thermodynamic irreversibilities and between logical and thermodynamic states, which is reminiscent of the  classic Gibbs' paradox about the joining of two volumes of the same gas; and 2)~from two unnecessary constraints imposed on the erase procedure.
Clarifying these points permits: to dissociate the notions of logical and thermodynamic irreversibilities; to
invalidate Landauer's result as being a general physical principle;
and to open the door to hardware implementations allowing erasure to follow a thermodynamically reversible, or at least quasistatic, path.

\end{abstract}
	
	\maketitle
	
	\section*{Introduction}

	In 1961 in a seminal paper\cite{Landauer_1961}, Rolf Landauer presented arguments which aim to set the physical basis of the relationship between information and thermodynamics. Based on the analysis of the behavior of a single particle in a bistable potential, which position figures a bit-state~0 or 1, he put forward the strong claim that erasing one bit of information dissipates at least $T\ln2$ of heat to the surroundings.
This statement makes any logically irreversible operation thermodynamically irreversible. Its strength is twofold: practical insofar as it would put a physical limit on computing power; and fundamental insofar as it would establish the link between information and energy which has fascinated people since Maxwell\,\cite{Maxwell_1872}, but this time independently to the Shannon's information theory\,\cite{Shannon_1948}. Landauer's result is now commonly regarded as a fundamental principle of physics\,\cite{Plenio_2001, Bormashenko_2019b, Herrera2020, Georgescu2021, VanVu2022, Bormashenko2024}.

The problem is that the Landauer's result, although it probably applies to many peculiar cases explaining why it has been confirmed experimentally\,\cite{Berut_2012, Berut_2015, Yan_2018, Proesmans_2020, Giorgini_2023, Binder_2023, Oriols_2023}, cannot be general for the simple reason that 
the logical irreversibility is a property in relation with a loss of information between the initial and final states of a system having undergone an operation, whereas we know since Clausius that the thermodynamic irreversibility is a property of the path that is used to make this operation.
It is up to our imagination to find the most economical.
Actually, the failure of the Landauer's result to become a general principle of physics has been already pointed out by several authors\,\cite{Earman_1999, Shenker_2000,Maroney2005, Norton2005}.
Perhaps these objections have not been well understood, but in any case certainly not convincing enough, because they patently had no effect.

In recent previous papers\,\cite{Lairez_2023, Lairez_2024a}, counterexamples have been  given for hardware implementations allowing to erase a bit in a quasistatic manner with an energy dissipation that tends towards 0 as the process is slowed down. But a  detailed comparative analysis of Landauer's implementation against these examples is lacking. 

The aim of this article is to dissect the Landauer's result and to show from where exactly it comes. First, it will be shown that it comes from a confusion between logical states and thermodynamic states that leads Landauer to link logical and thermodynamic irreversibilities.
A confusion which is not without recalling the second Gibbs' paradox\,\cite{Lairez_2024} about the reversibility of joining two volumes of gas  made of the same species: statistical mechanics considers the process to be irreversible, thermodynamics considers it to be reversible.
Actually, the two do not give the same meaning to the words \textquote{initial state} and do not treat the same information. Once this point understood the Gibbs' paradox appears to be veridical\,\cite{Quine1976}: the two points of view do not contradict each other.
The same goes for logical and thermodynamic irreversibilities, the two cannot be judged on the basis of the same information, so that, they are not necessarily linked.
The second point raised in this article is that the Landauer's claim also comes from two unnecessary constraints imposed on the erase procedure: that of being unique and that of following a mandatory scenario.
It will be shown that by clarifying this confusion or by releasing one of these two constraints, we open the door to as many thermodynamically reversible ways of erasure and to as many counterexamples to Landauer's result.
In doing so, it becomes excluded that Landauer's claim is a general principle.

\section{Landauer's arguments in detail}

\subsection{Minimum requirements for a hardware-bit}\label{mini}

Consider a physical system that is a candidate for the hardware implementation of a logical bit.
In order to be compatible with the encoding and processing of information, this system must meet certain requirements and exhibit certain properties.
The first requirement is linked to the choice of a binary encoding which imposes that:
\begin{enumerate}[label=\arabic*),
	leftmargin=*]\setcounter{enumi}{0}
	\item\label{first} {The system must present two distinguishable configurations, namely two logical states.}
\end{enumerate}

The second requirement is that the system must be able to maintain a given configuration, for a certain minimum time lapse, when the external conditions are stationary (at least the time needed to perform a calculation with the corresponding binary value).
So that:
\begin{enumerate}[label=\arabic*),
	leftmargin=*]\setcounter{enumi}{1}
	\item\label{second} {Configurations of the system must be lasting.}
\end{enumerate}

Obviously the system must be able to change configuration. But due to the persistence requirement, this can only be done by a set of external interventions, % actions (\textquote{action} in the common sense of the term, that is to say the fact of doing something), 
namely a procedure, imposed on the system by its surroundings.
For the storage of information, this procedure must be able to place the system in a particular configuration, for example the one corresponding to the value~0. This is the erase procedure (or erasure).
\begin{enumerate}[label=\arabic*),
	leftmargin=*]\setcounter{enumi}{2}
	\item\label{third} {The system must allow the existence of a set of external interventions allowing it to be set in the logical state~0, namely an erase procedure.}
\end{enumerate}

These three points constitute the minimum set of requirements that a hardware implementation of a bit must meet, and on which everyone can agree.
In addition, there is a statement regarding the functioning of the erase procedure itself.
\begin{enumerate}[label=\arabic*),
	leftmargin=*]\setcounter{enumi}{3}
	\item\label{fourth} {The erasure must have no side effect: only the bit to be erased is changed and nothing else.}
\end{enumerate}
This obligation is usually admitted in the literature. It is also fully consistent with the Landauer's reasoning. So, even if its necessity is not really clear it will not be examined in this paper, our argument should suffice as it~is.

\subsection{Landauer add-ons and result}

To the above constraints imposed on the hardware, 
Landauer\,\cite{Landauer_1961} followed by Bennett\,\cite{Bennett_1982} make some changes and add their own.
In this section we will not challenge them but will simply show how they lead to Landauer's famous result.

The first clear shift concerns what is considered as a valid logical state \myquote{which can hold information, without dissipation}\,\cite{Landauer_1961}.
This being admitted, a binary device consisting of a particle in a bistable potential is considered a generic hardware, probably the  simplest, which meets the first two requirements listed in the previous section. %: that of having two possible lasting logical states or configurations.
The next step concerns the nature of the potential in question. 
It is a thermodynamic potential, that is to say the free energy (or Helmholtz free energy) if the temperature is constant.
This is not explicit in Landauer's paper but nevertheless clear as soon as one talks about probability density, phase space and entropy of the particle\,\cite{Bennett_1982}.
Configurations where this potential is at minimum are thus assimilated to thermodynamic states. Additionally, in these two states the system is supposed to have the same internal energy and the same entropy (therefore the same free energy).
Hence, the first two constraints imposed to the hardware merge and become:
\begin{enumerate}[label=\roman*), leftmargin=*]\setcounter{enumi}{0}
	\item \textsc{One-to-one mapping}: The system must present two distinct logical states to which correspond bijectively two thermodynamic states of same internal energy and same entropy
\end{enumerate}

Landauer's other changes to what is required for a hardware-bit concern the erase procedure.
Landauer observes that 
\myquote{in most instances a computer pushes information around in a manner that is independent of the exact data which are being handled}\,\cite{Landauer_1961}. 
For the erase procedure, this is supported by the idea that it must be able to work for known or unknown input data.
Added to this the desire to avoid side effects in functional programming. So that an unknown input data to be erased must remain unknown and not copied somewhere. It follows that, according to Landauer, the procedure must be able to operate without needing to read the input data (it would then become known), thus preventing the use of a conditional statement. Hence the strong constraint imposed to the procedure:

\begin{enumerate}[label=\roman*), leftmargin=*]\setcounter{enumi}{1}
	\item\label{sec} \textsc{Uniqueness}: the erase procedure must be unique regardless of the input value and must avoid a conditional statement
\end{enumerate}

\begin{figure}[!htbp]
	\begin{center}
		\includegraphics[width=0.75\linewidth]{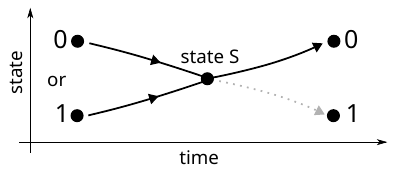}
		\caption{Erasure is supposed to bring the system into state~0, whether it was initially in state~0 or in state~1. Hence a junction of two paths (or a bifurcation if one intends to do a \textquote{reverse erasure}) that splits the process into two steps.}
		\label{junction}
	\end{center}
\end{figure}

Consider now the particle in a bistable potential landscape with two minima separated by an energy barrier. 
From the beginning of his paper\,\cite{Landauer_1961}, Landauer asks
whether we can construct a single time-varying force,
$F(t)$, which when applied to the particle will cause it to end up in the state~0, if it was initially in either the state~0 or the state~1 (see Figure \ref{junction}).
Note that in doing so, he implicitly makes a small shift in the meaning of  \textquote{external intervention} needed for the erase procedure, that here becomes \textquote{force applied to the particle}.
Next, Landauer observes that if this single time-varying force exists, then its time reversal must allow the trajectory of the particle to bifurcate, which is impossible with deterministic mechanics. It follows that $F(t)$ does not exist.
The erase process is thus necessarily uncontrolled and spontaneous somewhere, between a well determined position 0 or 1 and the junction point (or bifurcation in the other direction) where the system is neither in 0 nor in 1. Bennett\,\cite{Bennett_1982} calls the state of the system at this junction point the standard state or state~S.

The next crucial point is that for the system to evolve spontaneously from state~0 or 1 to state~S, and then at that point to be driven to state~0, the phase space volume of state~S must encompass at least those of state~0 and state~1. So that if these are equal, the volume of state~S is at least double that of the other two.
According to Landauer the erasing scenario is thus necessarily the following:

\begin{enumerate}[label=\roman*), leftmargin=*]\setcounter{enumi}{2}
	\item\label{sec} \textsc{mandatory erasing scenario}: 
\begin{itemize}[leftmargin=*]
	\item The erasure divides into two steps.%: 1)~from state 0 or 1 to state S; 2)~from state S to state 0.
	\item The first step is an expansion.\\ The second is a compression by the same factor.
	\item The expansion  of the phase space volume is at least by a factor of 2.
	\item  The expansion is free and cannot be quasistatic.
\end{itemize}
\end{enumerate}
The free expansion occurs without the hope of benefiting from it under the form of work provided to the surroundings. Exactly as the free expansion of a gas\,\cite{Bennett_1982}.
As for the compression step, it cannot be spontaneous and requires to be externally driven allowing it to be quasistatic.
This requires work from the surroundings that at least compensates for the increase in entropy experienced by the system during its expansion, that is to say at least $T\ln2$ (where the temperature $T$ is in Joule).
Work and heat are complementary quantities of the internal energy, thus it follows from the net energy balance of the two stages that the quantity $Q_\textsc{erase}$ of heat dissipated by the hardware-bit into the surroundings is such~as:
\begin{equation}
	Q_\textsc{erase} \ge T\ln2
\end{equation}
Hence the Landauer's limit.

\begin{figure}[!htbp]
	\begin{center}
		\includegraphics[width=1\linewidth]{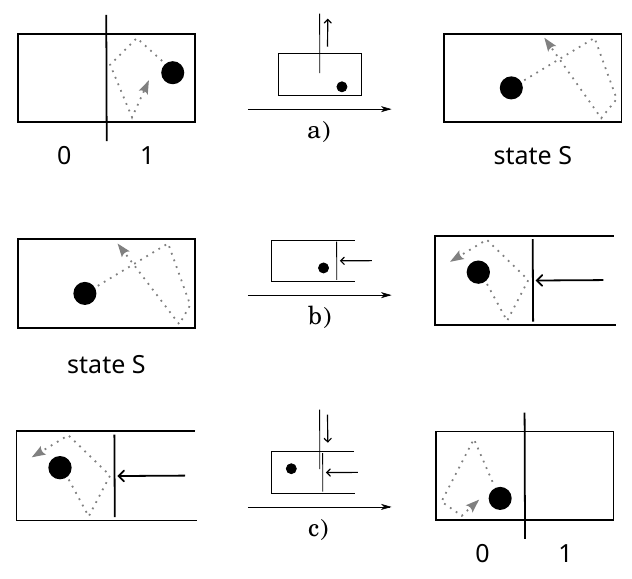}
		\caption{Landauer's erasure scenario visualized with a particle in a partitioned box\,\cite{Maroney2005}: a) remove the partition (free expansion); b) compress to state 0; c) put back the partition. Due to the constraint of uniqueness of the procedure, in stage a) a piston cannot be used to control the expansion because its side would depend on the initial logical state.}
		\label{maroney}
	\end{center}
\end{figure}

A clear representation of the Landauer's erasure scenario has been proposed by Maroney\,\cite{Maroney2005}. A particle is in a box with a vertical partition. 
If the particle is on the left, the logical value is 0, if it is on the right, it is 1. The Landauer's erase procedure is the following (see Figure \ref{maroney}):
\begin{enumerate}[label=\alph*)]
	\item Remove the partition (free expansion)
	\item Use a piston to push the particle on the left side (reversible compression)
	\item Put back the partition and remove the piston
\end{enumerate}

This representation of Landauer's erasure scenario is equivalent to that of a particle in a bistable potential well proposed by Bennett (\,\cite{Bennett_1982}, fig. 16).

\section{Irreversibilities}

At the root of Landauer's result is a preconceived idea:
\myquote{Logical irreversibility, we believe, in turn implies physical irreversibility, and the latter is accompanied by dissipative effects}, R. Landauer\,\cite{Landauer_1961}. In fact, it is not.
Logical irreversibility and thermodynamic irreversibility are both irreversibilities. It therefore seems natural to link the two, but this is typically a language trap. They have nothing to do with each other for two reasons: 1)~the former is a two-states properties whereas the latter is a path property; 2)~logical and thermodynamical states are not necessarily defined by the same information.

\subsection{States versus path properties}

Consider unary mathematical functions, those that take only one argument. Some, for example \textsc{increment}, \textsc{negation} etc., are such that from the image of these functions (the output) it is always possible to deduce the argument they had (the input). They are injections (or one-to-one functions).
These functions, when applied to a random source, preserve the Shannon's quantity of information emitted and are said to be logically reversible (note that the common sense of the term \textquote{information} is also perfectly suitable for the reasoning).
Others, when their output does not unequivocally define their input\,\cite{Landauer_1961, Landauer_1991}, do not preserve information and are said to be logically irreversible,
for instance \textsc{absolute value}, \textsc{square} etc.
From the knowledge of the output of these functions, we have no way of knowing what the input was.
For a Boolean function (a function that takes a binary value as an argument), classifying it as logically reversible or not can be done by examining only its truth table (see Table \ref{tab1}). Its logical irreversibility is an intrinsic property that only depends on its input and output. 

\begin{table}\label{tab1}
	\begin{tabular}{ccc}
		$
		\begin{array}{c|c}
			a &  \textsc{f}(a)
			\\ \midrule
			0 & 1\\
			1 & 0
		\end{array}
		$
		&~&
		$
		\begin{array}{c|c}
			a &  \textsc{g}(a)
			\\ \midrule
			0 & 0\\
			1 & 0
		\end{array}
		$
	\end{tabular}
	\caption{Truth tables for two mathematical Boolean functions: \textsc{f} is logically reversible and \textsc{g} logically irreversible.}
\end{table}

Although these definitions are mathematically clear, their transposition to computer science requires some clarification.
%In mathematics, whether or not a function preserves information only depends on its truth table.
Once implemented in a computer, whether or not a function preserves information depends not only on its truth table, but also on whether or not the input data has been copied somewhere (including into our brain's memory) before or during the function's execution.
If so, there is no loss of information and all functions are always logically reversible.
We can make as many copies of a piece of data as we want, the amount of information is the same: information cannot be given twice. Conversely, we can erase many copies and retain the information. It is only by erasing the last one that we lose information.
It follows that \textit{a priori} the erasure of a bit falls into two cases:
\begin{itemize}[leftmargin=*]
	\item Erase a bit of uncopied or original data which is logically irreversible and results in loss of information;
	\item Erase a bit of copied data which is logically reversible and leaves the information unchanged.
\end{itemize}
However, whatever the case that is considered, the logical irreversibility of erasing one physical bit only depends on the initial and final states of the overall system, that is to say the hardware-bit plus its surroundings (where an eventual copy of the data is located). The logical irreversibility remains a property of the initial and final states of the overall system.

In thermodynamics, nothing is reversible \textit{stricto sensu}. However, some processes can be decomposed into a series of infinitesimal changes which rate of succession can be entirely externally controlled. In this case, the process can be driven forwards or backwards with a path-coincidence that increases with decreasing rate. So that for one cycle, the net energy balance is as small as desired, with no finite limit when the process is driven more and more slowly. A typical example is that of the monothermal expansion-compression cycle of a gas with a piston.
These processes are called quasistatic, but often also reversible for convenience of language.
Thermodynamically irreversible processes are those that cannot be externally controlled in this way. They occur spontaneously in an internally driven manner. According to the second law of thermodynamics for an isolated system, thermodynamically irreversible processes necessarily leads to a state of higher entropy.
But the character of irreversibility is not inherent to the initial and final states. It is inherent to the path used to connect these two states.
A typical example is that of the adiabatic free expansion of a gas. 
The process is irreversible, but the system can return to the initial state by the mean of a reversible isothermal compression. The latter path having also the ability of being used in both directions.
In short:
\begin{itemize}[leftmargin=*]
	\item Logical irreversibility is a property of the initial and final states of the overall system in relation with a loss of information: \\	
	We cannot return the system to its initial logical state because we have no way of knowing where it came from.
	\item Thermodynamic irreversibility is a property of a particular path that connects two states:\\	 
	We cannot return the system to its initial thermodynamic state by the reverse path because it is impracticable in that direction, but other paths may exist.
\end{itemize}

Because of this fundamental difference, there is absolutely no reason why the two notions should be linked. To argue otherwise leads to serious inconsistencies. 

\subsection{Independence of the two}\label{sec_theorem}

The fact that copying data before the execution of all instructions in a computer program makes any computation logically reversible is clear and usually acknowledged:
\myquote{It is easy to render any computer reversible in a rather trivial sense, by saving all the information it would otherwise have thrown away}, C. Bennett\,\cite{Bennett_1982}.
\myquote{Computation that preserves information at every step along the way (and not just by trivially storing the initial data) is called reversible computation}, R. Landauer\,\cite{Landauer_1991}.

So now, let us consider the following thought experiment. 
Consider
a given hardware-bit of original (uncopied) data. Make a physical copy of it. 
Separate the two hardware so that they are physically independent (non interacting).
Erase the first, the operation is logically reversible. Erase the second, the operation is logically irreversible.
The Clausius entropy (that of thermodynamics) is a state quantity
and its variation $\Delta S$ is the same for the two erased systems since they have the same initial and final thermodynamic states.
In thermodynamics the only criterion to decide for the irreversibility of a process is the Clausius inequality that writes at constant temperature $T$ (in Joule):
\begin{equation}\label{Clausius_ineq}
	-Q\ge -T\Delta S
\end{equation}
where $-Q$ is the quantity of heat dissipated by the system (and received by the surroundings). The equality holds for thermodynamically reversible processes and the inequality for irreversible ones.
In the case of our two hardware-bits (the original and its copy), they are identical and physically independent, so there is no obstacle to using exactly the same process to erase each of them. Doing so they dissipate exactly the same quantity of heat. Therefore, the difference between $T\Delta S$ and $Q$ for the two erasures is the same as well as their thermodynamic reversibility, although their logical reversibility differs. Hence the unavoidable theorem:

\begin{thm}[of independence of reversibilities]
	The thermodynamic reversibility of a process is the same regardless of its logical reversibility.
\end{thm}

Consider a hardware-bit subjected to a thermodynamic process that has the property of being reversible or not and possibly logically reversible or not. 
In principle, we have four possibilities for combining these properties.
The above theorem tells us that none of these combinations are impossible, but not that examples are easy to conceive.

\subsection{Constraint of \textsc{one-to-one mapping}}

Landauer's preconception about the link between the two types of irreversibilities forces him to equate logical and thermodynamic states. In Landauer's mind, the two are bijectively related and coincide. This idea is not correct.

Consider a coin with two identical sides. When thrown on the table, the coin has only one stable thermodynamic state with two compatible microstates (one per side). But as is, the coin cannot serve for binary encoding. For this, the two sides must be able to be distinguished. For instance, one of the two faces can be marked with a cross. Having done this, the thermodynamics of the system is unchanged.
We still have only one thermodynamic state but two logical states that map to the microstates. Here, the one-to-one correspondence applies between logical values and microstates, not between logical values and thermodynamical macroscopic state.
This is not just a question of vocabulary, thermodynamics laws apply to macroscopic states not to microstates. 
The two are not appreciated based on the same information.
Thinking otherwise leads to the famous second Gibbs' paradox about the joining of two volumes of the same gas\,\cite{Lairez_Stirling, Lairez_2024}.

The second Gibbs' paradox comes from an apparent contradiction between thermodynamics and statistical mechanics\,\cite{Lairez_Stirling, Lairez_2024}. Consider a gas put in a box. Divide the box into two volumes at same temperature and pressure by putting a partition.
In thermodynamics, the removal and the replacement of the partition is not accompanied by any exchange of heat or work with the surroundings, so that the two states (joined or disjoined volumes) have the same Clausius entropy. The removal of the partition is thermodynamically reversible.
In statistical mechanics, a direct (and naive) application of Boltzmann's equation for the entropy leads to the conclusion that the latter increases with the removal of the partition (the phase space volume per particle increases). So that the system cannot be returned to its initial state without an energy expenditure, just by replacing the partition.  The removal of the partition is irreversible.
Hence the paradox. Who is telling the truth? Actually here, 
statistical mechanics and thermodynamics do not give the same meaning to the word \textquote{initial state}. 
The first considers the initial state as being defined by a well determined number of particles in each compartment and also that particles are traceable and must return to their original compartment for the initial state to be restored. The second does not care about these two pieces of information.
Interestingly, between these two extreme conceptions of what \textquote{initial state} means, all intermediate possibilities exist (see \cite{Lairez_2024} sec. 3.1.3) depending on the degree of incompleteness of our knowledge about this initial state. But in any cases, thermodynamics always judges a \textquote{state} on the basis of an incomplete knowledge that is a subset of the overall information. 150 years ago Gibbs wrote:
\myquote{It is to states of systems thus incompletely defined that the problems of thermodynamics relate}\,\cite{Gibbs1874} p.228.
This point is essential to untangle logical and thermodynamic irreversibilities.

Let us propose a \textquote{hardware-bit version} of the second Gibbs' paradox.
Consider the memory of a computer supposed to work in a continual dynamic context, that is to say it continually receives new original and random data of unknown distribution and does calculation with~it.

\begin{itemize}[leftmargin=*]
	\item \textbf{Question}: Can the same bit be erased and reused cyclically?
\end{itemize}
\noindent 
The answer depends on our ability to make the bit pass through the same initial state again (otherwise it is not a cycle). Thus, two statements can be made:
\begin{itemize}[leftmargin=*]
	\item \textbf{Statement A}: We are able to put the bit back to the same initial state.
	\item \textbf{Statement B}: The computer cycle is logically irreversible, so that the bit cannot be returned to an initial state of which we do not know what it was.
\end{itemize}

The two statements contradict each other in a very similar manner as in the Gibbs' paradox exposed previously.
Which is the correct statement?
The common sense and the experience tell us that the bit can be reused so that A is correct.
But how can we rule out statement~B? The only solution (that is also that of the Gibbs' paradox) is that the two statements are correct but do not give the same meaning to the word \textquote{initial state}. Statement A means \textquote{thermodynamic state}, whereas statement B means \textquote{logical state}. The initial thermodynamic state of the bit is actually that of a bit which has a unknown, or unpredictable, logical state.
The two are different.
They can be mapped by a one-to-one correspondence in some circumstances, but generalizing this correspondence would lead to inconsistencies.

Actually in the framework of Landauer's hardware and erasure (Figure \ref{maroney}) it is possible to obtain for the erasure all the possible combinations discussed at the end of sub-section \ref{sec_theorem}: logically reversible or not and thermodynamically reversible or not.
For this, the bit is materialized under the form of a particle in a box partitioned in two volumes.
For the sake of simplicity, the operation undergone by the bit is reduced to the first step of Landauer's erasure, the one which corresponds to the expansion of the phase space and is responsible for Landauer's result. Here, this step is performed by removing the partition between the two compartments of the box (step a, Figure \ref{maroney}). 
For commodity of language, let us call this operation \textquote{erasure} (with quote because it is not really speaking an erasure i.e. \textsc{set to 0}) and that of putting the partition back \textquote{reverse erasure}.
The \textquote{erasure} operation causes the final state of the bit to be the same whatever its initial logical state, so its logical irreversibility is judged solely on the basis of the existence or not of a copy.
The hardware-bit in question belongs to an overall system (the universe) made partly of two computers. 
The first computer can be viewed as a demon, he knows everything about the present moment of the universe, so that he has the ability to tell whether of not the \textquote{erasure} causes a loss of information i.e. whether or not it is logically irreversible, that is, in our case whether the initial value of the bit has been copied somewhere or not.
The second computer is an observer, which can be identified with ourselves. The only things the observer knows about the universe are those recorded in his brain's memory (a subset of the universe's memory).
Regarding the \textquote{erasure} and \textquote{reverse erasure} sequence, the only ability of the observer is to tell whether or not it restores the hardware-bit to its initial thermodynamic state based on what he knows about it, 
i.e. whether or not the \textquote{erasure} is thermodynamically reversible.

In short, logical reversibility of \textquote{erasure} is judged by the demon based on the total information. 
Whereas thermodynamic reversibility is judged by the observer, according to its incomplete knowledge, that is, a partial information.

\begin{figure}[!htbp]
	\begin{center}
		\includegraphics[width=0.8\linewidth]{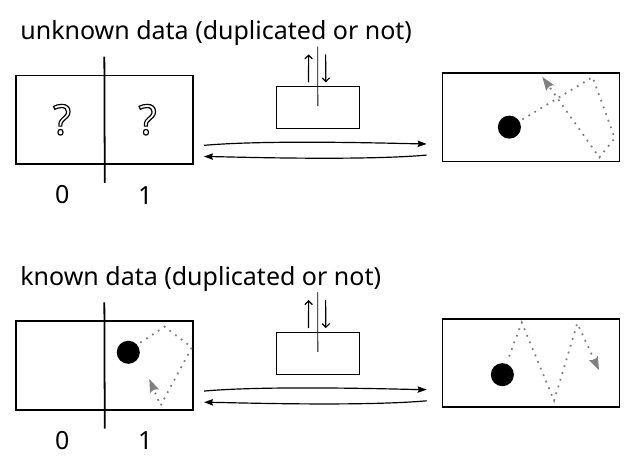}
		\caption{Hardware-bit under the form of a particle in a partitioned box. The operation of removing the partition is judged as being logically reversible or not depending on whether a copy of the initial bit-value exists or not. 
		But, it is judged thermodynamically reversible or not depending on whether the return of the partition restores or not the initial state on the basis of our knowledge of what it was.}
		\label{unknown}
	\end{center}
\end{figure}

\begin{figure}[!htbp]
	\begin{center}
		\includegraphics[width=1\linewidth]{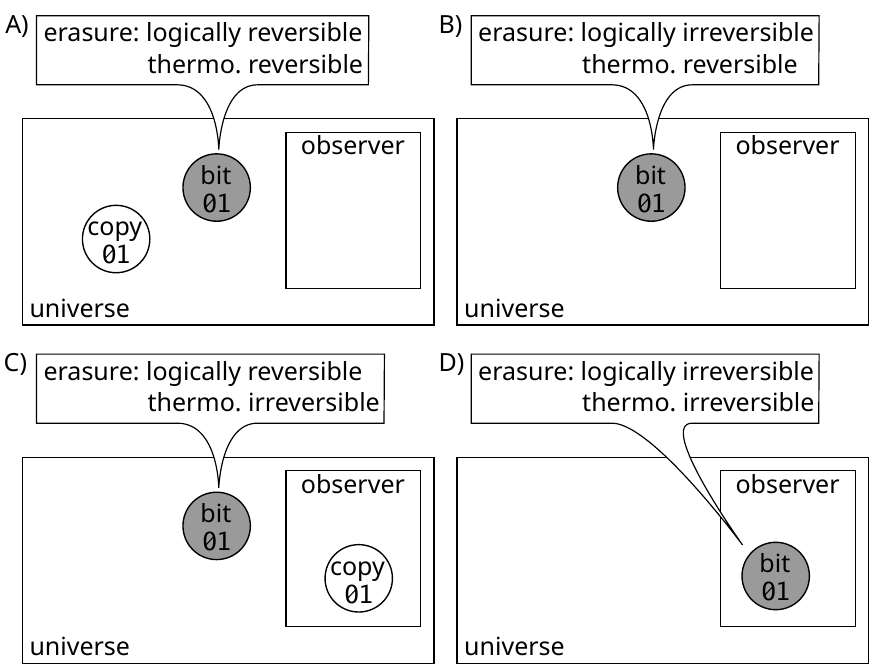}
		\caption{Four possibilities to erase a bit (see text): depending on whether the initial bit-value has been duplicated or not; and whether it is known to the observer or not.}
		\label{info}
	\end{center}
\end{figure}

We have four possible combinations of an initial bit-value, known or unknown by the observer, and duplicated somewhere or not (see Figure \ref{unknown} and \ref{info}):
\begin{enumerate}[label=\Alph*),leftmargin=*]
	\item The bit-value is duplicated but unknown\\ (the copy is not in the observer's memory): 
	\begin{itemize}
		\item The operation is logically reversible.
		\item The operation is thermodynamically reversible: after a cycle the logical state is unknown but it was also unknown before, so based on the observer's knowledge the initial state is restored.
	\end{itemize}
	\item The bit-value is not duplicated and is unknown\\ (no copy anywhere):
	\begin{itemize}
		\item The operation is logically irreversible.
		\item The operation is thermodynamically reversible for the same reason as in the previous case
	\end{itemize}
	\item The bit-value is duplicated and known\\ (there is a copy which (or the original, or both) is in the observer's memory):
	\begin{itemize}
		\item The operation is logically reversible.
		\item The operation is thermodynamically irreversible: based on the observer's knowledge, the return of the partition is insufficient to systematically restore the initial state.	
	\end{itemize}
	\item The bit-value is not duplicated but known\\ (the original bit is in the observer's memory):
	\begin{itemize}
		\item The operation is logically irreversible.
		\item The operation is thermodynamically irreversible for the same reason as in the previous case.
	\end{itemize}
\end{enumerate}

One find the four possibilities expected from the theorem of independence of reversibilities (sub-section \ref{sec_theorem}).
The second case was first proposed by Maroney\,\cite{Maroney2005} as a counterexample to Landauer's claim.
But the author has limited its applicability to the case where, although the binary-value is unknown, its probability distribution is known. So that the position of the partition can be such that the ratio of the two sub-volumes thus separated is equal to the inverse of that of the two probabilities.
In fact, this restriction is unnecessary.
In case the distribution is totally unknown,  we have no reasonable choice but to assume that they are uniformly distributed. There is no reason to suppose otherwise. This is the Laplace's \textquote{principle of insufficient reason}, renamed in statistical mechanics \textquote{the fundamental postulate} (\cite{Balian_1991} p.143). Its more general version, known as \textquote{maximal entropy principle}, leads to the same result.

\begin{figure}[!htbp]
	\begin{center}
		\includegraphics[width=1\linewidth]{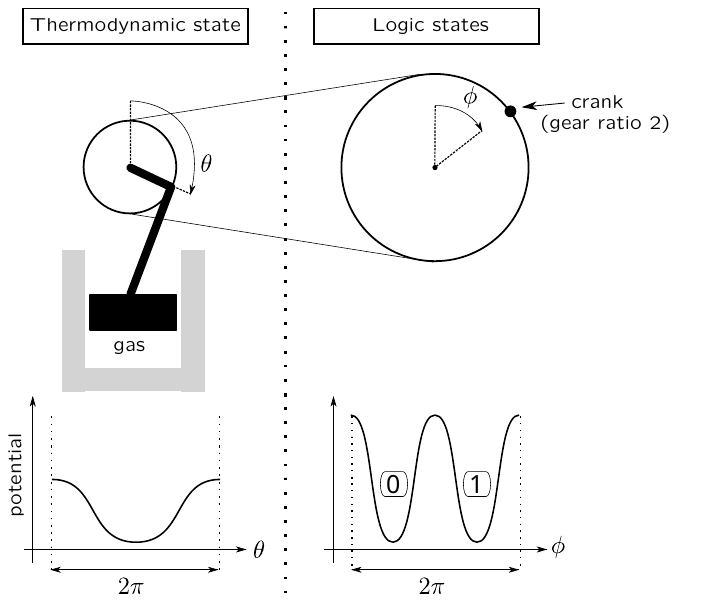}
		\caption{Two-to-one correspondence between logical states (right) and thermodynamic state (left). The latter is stable when the gas under the piston is at the same pressure as above. The piston is driven by a crank and a gear ratio of 2, so that the only stable position of the piston corresponds to two stable positions for the crank which materialize the binary values.
		The logical state is evaluated based on the position of the crank, whereas the thermodynamic state is determined on the basis of the partial information concerning the position of the piston.}
		\label{fig_two2one}
	\end{center}
\end{figure}

In the previous example, the logical and thermodynamic states, as well as the corresponding reversibilities, are totally uncoupled. But this may seem a bit artificial and specific to the hardware implementation that was chosen. In fact, it is Landauer's own, but one could consider that with this reasoning based only on information, what is missing and unclear is whether (independently of logical irreversibility) erasure can be quasistatic (i.e. controlled and as slow as desired).
In other words, can releasing the constraint of one-to-one mapping between logic and thermodynamics allow an implementation somewhat similar to that of the coin, but allowing the erasure to be quasistatic?
Such an implementation has been proposed in ref.\,\cite{Lairez_2023}.
 In this example, the uncoupling of logic and thermodynamics is ensured by the mean of a mechanical frequency divider: a change in the value of the logical bit, i.e. half of a logical cycle, results in a complete thermodynamic cycle (see Figure \ref{fig_two2one}). 
In the terms of \textquote{demon and observer} previously used, here the demon judges the logical state based on the position of the crank. Whereas the observer judges the thermodynamic state on the basis of the partial information concerning the position of the piston.
With this hardware the Landauer's constraint of uniqueness can be satisfied. The erasing scenario is also the same as Landauer's, except that this time during the expansion no uncontrolled leakage can occur between two thermodynamic potential wells because there is only one! Actually, state~S can be reached without moving the piston by continuously decreasing the gear ratio\,\cite{Lairez_2023} until the energy barrier between states 0 and 1 is smaller than the thermal energy.
So that the erase procedure is guaranteed to be thermodynamically reversible.

\section{Erasure}

As explained in the previous section, logical and thermodynamic states do not necessarily correspond one-to-one. This is therefore sufficient in itself to provide counterexamples to Landauer's claim.
But what happens if the two coincide? Can erasure still be quasistatic? Are the two constraints imposed by Landauer, that of uniqueness and that of the erasure scenario, really necessary? And finally, does the releasing of these constraints allow for quasistatic erasure?

\subsection{Constraint of \textsc{uniqueness}}

The constraint of uniqueness of the erase procedure must be analyzed according to the obligation to avoid side effects, but also according 
to the two cases: known or unknown data, that is to say copied or not into the memory of the device in charge of the erasure.
So let us first see if these two cases are actually relevant.

This paper is devoted to the link between thermodynamic and logical irreversibilities.
Thermodynamic systems, as their name suggests, are dynamic. Even in the stable state of equilibrium, they continually change their microscopic configuration, i.e. their microstate. 
So, for the comparison with a bit to be relevant, the latter must also be considered in a dynamic context where its value is continually renewed by a computer operating indefinitely, such as the one supposed to replace Maxwell's demon\,\cite{Maxwell_1872} (or the single particle version of Szilard\,\cite{Szilard_1964}). In this context, if the eventual memory buffer in which data are copied has a finite size, the logical irreversibility of erase operations is just postponed until the buffer is filled. In other words, if the buffer is intended to be reused cyclically the logical irreversibility of erasing a bit is simply shift to that of erasing its copy, which itself has no copy.

At first sight, finite buffer memory seems to be the rule, for this reason one might argue that the case of copied data does not deserve consideration.
In fact, one can imagine a situation where the memory size is so large that the time needed to fill it is never reached.
This is exactly what happens for macroscopic thermodynamic systems that are theoretically expected to behave cyclically according to Poincaré's recurrence theorem, but in practice never do so because of their very large number of microstates\,\cite{Lairez_2024}. The period of recurrence is greater than the age of the universe.
Similarly, one can imagine a memory size such that it would take longer to fill than the lifetime of the computer.
It follows that the case of copied data is well worth considering, especially in a context where Landauer's principle is claimed to have major implications not only for computer science, but also for cosmology and the fundamental laws of the universe\,\cite{Landauer_1991}.

At first glance the constraint of uniqueness may seem legitimate for data which are unknown by the device ensuring the erasure, i.e. not copied into its memory (in fact it is not as we will see later) according to the argument that a conditional statement cannot be used if data is unknown. But this argument is not valid for the case of known data, for which a conditional statement can be used, with no side effect, that adapts the operation to the data to be erased. 
The rest of the reasoning is straightforward:
\begin{itemize}[leftmargin=*]
	\item if there exists a reversible compression path from state~S to state~0, by symmetry there is one from state~S to state~1 (the choice of \textsc{set to 0} to erase the bit is arbitrary, Landauer has chose \textsc{set to 1}\,\cite{Landauer_1961}); 
	\item if there is a reversible path from state~S to state~1, then it can be used in the other direction for a reversible expansion from state~1 to state~S.
\end{itemize}
The whole erasure can thus concretely be achieved with the sequence:
\begin{itemize}[leftmargin=*]
	\item \textsc{if} the initial logical state is 0 
	\textsc{then} do nothing;
	\item \textsc{else} do: 1)~the reversible expansion from state~1 to state~S; then 2)~the reversible compression from there to state~0.
\end{itemize}
Therefore, the whole erasure for known data can be  thermodynamically reversible.

Consider now the case of unknown data.
In a dynamic context, it is possible to use a single given bit to \textsc{read} (and \textsc{copy}) the input argument of the erasure without side effect. Actually, in a cyclic stationary functioning, the copy made in cycle $n$ will be erased by the new copy of the cycle $n+1$.
So the logically irreversibility of the erasure is only delayed by one cycle, but the procedure can now use a conditional statement and a thermodynamically reversible path.
There is absolutely no impediment to this, it is just less convenient than if the operation was unique (note that this point was already mentioned in \cite{Norton2005} p.400 but not really discussed).

\subsection{Constraint of the \textsc{scenario}}

Let us now examine the scenario that, according to Landauer, the erasure must necessarily follow once the two previous constraints are admitted.

At the root of this scenario is the preconceived idea of what exactly \textquote{external intervention} means in the third requirement that a hardware-bit must satisfy (section \ref{mini}, point \ref{third}). In Landauer's mind, this means a time-varying external force applied to the particle (see~\cite{Landauer_1961} p.184).
This interpretation is very restrictive. In thermodynamics \textquote{external intervention} means \textquote{energy exchange} and not only \textquote{mechanical work}, even \textquote{mechanical work} does not necessarily mean a force directly applied on the particle. Landauer himself in its erasure scenario involves external interventions that are not in the form of a force directly applied to the particle, the bias is, but not the lowering and raising of the energy barrier (or the removal and return of the partition in Figure \ref{maroney}).
Moving away from this restrictive interpretation allows for new scenarios.

To fix ideas, let us first consider a particle in a potential energy landscape, which can be viewed as a topographic relief, that is to say the result of: 1)~a short range repulsive interaction between the particle and a non-flat surface; and 2)~an attractive field $E$ perpendicular (in average) to the surface. The force $F$ experienced by the particle does not only result from the field, but also from its interaction with the surface. So the force depends on where the particle is located and can be different depending on whether it is in position~0 or~1.
It immediately follows that a single time-varying field $E(t)$ leads to different time-varying forces $F(t)$ according to the (initial) position of the particle, thus reconciling the requirement of uniqueness with deterministic mechanics. 
When doing the first step of erasure, that is between state 0 or 1 to state S (see Figure~\ref{junction}), it is as if a conditional statement was carried by the pre-existing topographic relief and did not need to be implemented by the erase procedure itself. When doing the second step, at the bifurcation from state S to either state 0 (full black line in Figure \ref{junction}) or 1 (doted gray line), a bias can be applied to choose the path leading to 0, exactly as with the Landauer's procedure.
This observation completely invalidates the Landauer's reasoning that led him to the conclusion that the first step of the erasure is necessarily spontaneous and cannot be quasistatic. In fact, this is not necessary at all.

\begin{figure}[!htbp]
	\begin{center}
		\includegraphics[width=1\linewidth]{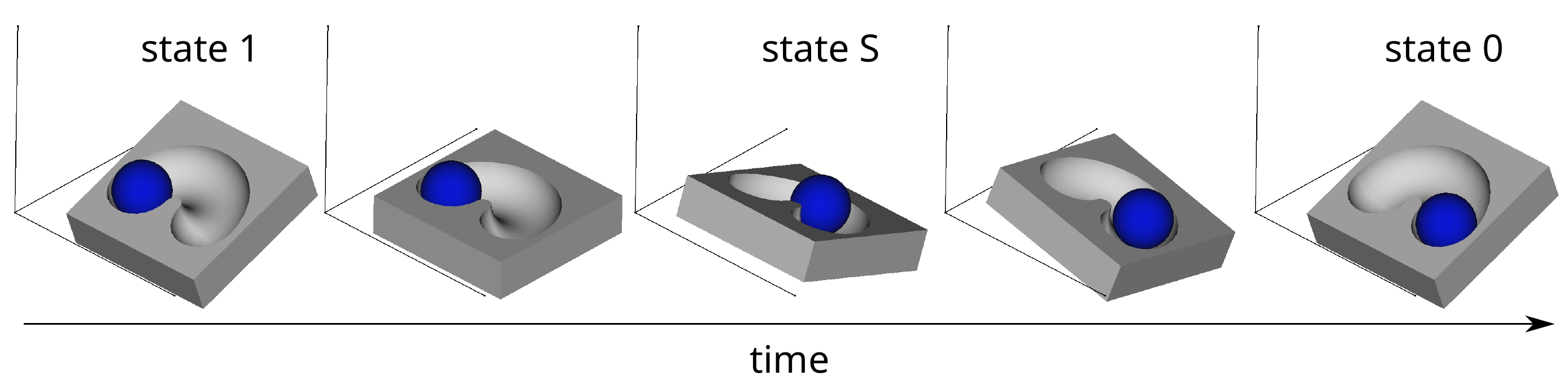}
		\caption{Hardware-bit under the form of a particle in a topographic relief, i.e. a short range repulsive surface plus a vertical attractive field. A unique appropriate sequence of rotation of the surface permits the erasure regardless of the initial state. The force experienced by the particle can be as small as desired, so that due to friction the motion of the particle can also be as slow as desired and the process is quasistatic.}
		\label{field}
	\end{center}
\end{figure}

In Figure \ref{field} an example of hardware implementation based on what has just been said is given. Here the field is vertical (we can think of it as gravity) and a unique appropriate sequence of rotation of the surface allows erasure of the bit regardless of its initial state (another implementation might involve a rotating field).
The particle follows the opposite direction of the maximum elevation gradient (maximum potential gradient) and experiences a force in this direction proportional to the  angle of inclination (for small values), so that it can be as small as desired. If the motion is subject to friction, it can also be as slow as desired depending on the inclination and the process can be quasistatic.

\section{Conclusion}

Landauer's result has been obtained in a particular context, where logic and thermodynamic states and levels of information coincide, and where some constraints are imposed to the erase procedure. In this particular context, there is absolutely no flaw in Landauer's reasoning. So that it is not surprising that it has been confirmed experimentally\,\cite{Berut_2012, Berut_2015, Yan_2018, Proesmans_2020, Giorgini_2023, Binder_2023, Oriols_2023}.
But that is not the main point. It would rather be to know whether or not this result can be a general principle of physics. Clearly not, whatever the side one approaches the problem from.
Logic and thermodynamic coincidence, as well as erasure constraints are certainly more convenient from a practical point of view. Hardware implementations that follows these specifications are probably more "computer-compatible", or more efficient for computation than others. But a fundamental law of physics cannot be based on a convenience argument.
Nature has no obligation to be convenient to us.

\bibliography{~/Documents/Articles/weri_biblio.bib}

\end{document}